\documentstyle[prl,aps,epsf,multicol]{revtex}
\newcommand{\be}{\begin{equation}}
\newcommand{\ee}{\end{equation}}
\newcommand{\bea}{\begin{eqnarray}}
\newcommand{\eea}{\end{eqnarray}}
\begin{document}

 \title{Exact Tagged Particle Correlations in the Random Average Process}
 \author{R. Rajesh$^1$ and Satya N. Majumdar$^{1,2}$}
 \address{ {\small 1. Department of Theoretical Physics, Tata Institute of
Fundamental Research, Homi Bhabha Road, Mumbai 400005, India.}\\
 {\small 2. Laboratoire de Physique Quantique (UMR C5626 du CNRS),
Universit\'e Paul Sabatier, 31062 Toulouse Cedex, France.}}
 \date{\today}
 \maketitle
 \widetext
 %%%%%%%%%%%%%%%%%%%%%%%%%%%%%%%%%%%%%%%%%%%%%%%%%%%%%%%%%%%%%%%%%%%%%%%%    

\begin{abstract} 

We study analytically the correlations between the positions of tagged
particles in the random average process, an interacting particle system in
one dimension. We show that in the steady state the mean squared
auto-fluctuation of a tracer particle grows subdiffusively
$\sigma_0^2(t)\sim t^{1/2}$ for large time $t$ in the absence of external
bias but grows diffusively $\sigma_0^2(t)\sim t$ in the presence of a
nonzero bias. The prefactors of the subdiffusive and diffusive growths as
well as the universal scaling function describing the crossover between
them are computed exactly. We also compute $\sigma_r^2(t)$, the mean
squared fluctuation in the position difference of two tagged particles
separated by a fixed tag shift $r$ in the steady state and show that the
external bias has a dramatic effect in the time dependence of
$\sigma_r^2(t)$. For fixed $r$, $\sigma_r^2(t)$ increases monotonically
with $t$ in absence of bias but has a non-monotonic dependence on $t$ in
presence of bias. Similarities and differences with the simple exclusion
process are also discussed.

 \end{abstract}
%%%%%%%%%%%%%%%%%%%%%%%%%%%%%%%%%%%%%%%%%%%%%%%%%%%%%%%%%%%%%%%%%%%%%%%%%%
\begin{multicols}{2}

\section{Introduction}

Interacting particle systems in one dimension are amongst the simplest
examples of many body systems that are far from equilibrium\cite{liggett}.
One of the most studied examples is the simple exclusion process in one
dimension. In this system, each site of a one dimensional lattice is either
occupied by a hardcore particle or it is empty. In a small time interval $dt$,
each particle attempts to hop to the neighboring lattice site on the
right with probability $pdt$, to the left neighboring site with
probability $qdt$ and stays at the original site with probability
$1-(p+q)dt$. An attempted hop is completed provided the target site is
empty. A wealth of results are known for this system\cite{liggett,spohn,DE}.

Another interacting particle system in one dimension that has attracted
recent interest is the random average process (RAP)\cite{KG,RM1}. In the
RAP, particles are located on a real line as opposed to a lattice in the
simple exclusion process. Let $x_i(t)$ be the position of the $i$-th
particle at time $t$ (see Fig. 1).
\begin{figure}
\narrowtext\centerline{\epsfxsize\columnwidth \epsfbox{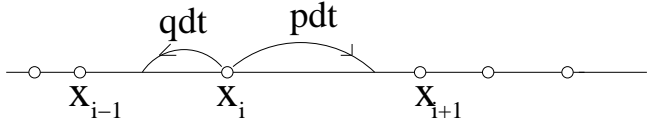}}   
\caption{The stochastic moves in the RAP.}
\end{figure}
\noindent
In a small time interval $dt$, each particle jumps to the right with
probability $pdt$ by an amount $r_i^{+}(x_{i+1}-x_i)$, to the left with
probability $qdt$ by an amount $r_i^{-}(x_i-x_{i-1})$ and stays at its
original location with probability $1-(p+q)dt$. Here $r_i^{+}$ and
$r_i^{-}$ are independent random variables drawn from the interval $[0,1]$
with identical probability density function (pdf) $f(r)$. Thus the jumps
in either direction is a random fraction of the gap to the nearest
particle in that direction. For convenience, we have defined the RAP with 
random sequential dynamics, though it has been studied with parallel
dynamics as well\cite{KG,RM1}. The detailed study of the RAP is important
since it has shown up either directly or in disguise in a variety of
problems including traffic models\cite{KG}, models of mass
transport\cite{RM1}, models of force fluctuation in bead packs\cite{SC},
models of voting systems\cite{melzak,FF}, models of wealth
distribution\cite{IKR} and the generalized Hammersley process\cite{AD}.
Like the simple exclusion process, some aspects of the RAP are
analytically tractable\cite{KG,RM1,ZS}. In this paper, we derive some new
exact results on the tracer fluctuations in the RAP where the dynamics of
tagged particles are followed.

The tracer diffusion has been studied in detail for the simple exclusion
process and many interesting results are known\cite{liggett}.  In the
exclusion process, the combined effect of hardcore interaction and the
external bias $(p-q)$ shows up rather dramatically in the asymptotic long
time behavior of the mean squared auto-fluctuation in the position of a
tracer particle in the steady state. If $\zeta_i(t)=x_i(t)-\langle
x_i(t)\rangle$ denotes the deviation in the position $x_i(t)$ of the
$i$-th particle from its average value, then the mean squared
auto-fluctuation is defined as, $\sigma_0^2(t_0,t_0+t)=\langle
[\zeta_i(t_0+t)-\zeta_i(t_0)]^2\rangle$, where $t_0$ is the waiting time
after which one starts measuring the fluctuations. In the steady state
$t_0\to \infty$, the asymptotic behavior of $\sigma_0^2(t)=\lim_{t_0\to
\infty}\sigma_0^2(t_0,t_0+t)$ for large $t$ is known\cite{liggett}. In
absence of external bias ($p=q=1/2$), i.e. for the symmetric exclusion
process (SEP), $\sigma_0^2(t)\sim A t^{1/2}$ for large $t$ where the
constant $A=(2/\pi)^{1/2}(1-\rho)/\rho$ is known exactly in terms of the
density $\rho$ of the particles\cite{harris,arratia,AP}. This slow
subdiffusive growth is due to the caging effect arising from hard core
exclusion in one dimension where a particle is always hemmed in by its
neighbors. However in the asymmetric case (ASEP) when a nonzero bias
$p-q>0$ is switched on, one finds, somewhat unexpectedly,
$\sigma_0^2(t)\sim D t$ for large $t$ where the tracer diffusion
coefficient $D=(p-q)(1-\rho)$\cite{DF,KvB}. The crossover from the
subdiffusive to diffusive behavior of $\sigma_0^2(t)$, as an infinitesimal
bias is switched on, was understood in a physically transparent way via a
rather unusual mapping of the exclusion process to a $(1+1)$-dimensional
interface model\cite{MB1,MB2}. This mapping also established that an
appropriately defined sliding tagged-particle correlation function varies
anomalously as $t^{2/3}$\cite{MB1}. This anomalous $t^{2/3}$ growth
also shows up in the mean square fluctuation of the center of mass of the
particles when viewed from a special moving frame\cite{vBKS}.

A question then arises naturally: what are the corresponding results on
the tracer diffusion for the RAP? The only known result is for the fully
asymmetric RAP with $q=0$ (and time rescaled by $p$) where the particles
move only to the right. In this limit, $\sigma_0^2(t)$ was
computed by Krug and Garcia using a phenomenological hydrodynamic Langevin
equation based on heuristic arguments as well as using an independent jump
approximation\cite{KG}. Their result shows that $\sigma_0^2(t)\sim
D_{1}t$ for large $t$ with $D_{1}={\rho}^{-2}\mu_1\mu_2/(\mu_1-\mu_2)$
where $\rho$ is the density of the particles and $\mu_k=\int_0^{1}dr r^k
f(r)$ is the $k$-th moment of the pdf $f(r)$. Later, Sch\"utz attempted to
derive this result rigorously\cite{schutz} by writing down the exact
equation of evolution of the equal time correlation function $G_r(t)=\langle
\zeta_0(t)\zeta_r(t)\rangle$ and then using a chain of arguments. Note
that the definition $\sigma_0^2(t_0,t_0+t)= \langle
[\zeta_i(t_0+t)-\zeta_i(t_0)]^2\rangle$ involves both the variance
$\langle \zeta_i^2(t)\rangle$ which is an equal time observable as well as
the unequal time correlation $\langle \zeta_i(t_0)\zeta_i(t_0+t)\rangle$.
Thus a proper approach, as followed in this paper, would be to compute
these correlation functions exactly and then take the steady state $t_0\to
\infty$ limit.

The main results of this paper can be summarized as follows:
 \begin{enumerate}
 \item We compute exactly the mean squared auto-fluctuation in the
displacement of a single tracer particle, $\sigma_0^2(t_0,t_0+t)=\langle
[\zeta_i(t_0+t)-\zeta_i(t_0)]^2\rangle$ for large $t_0$ and $t$ for all
values of $p$ and $q$ in the RAP.  In the steady state $t_0\to \infty$, we
show that $\sigma_0^2(t)=\lim_{t_0\to \infty}\sigma_0^2(t_0,t_0+t)\sim
A_{SRAP} t^{1/2}$ for large $t$ for the symmetric RAP (SRAP) with $p=q$.
For the asymmetric RAP (ARAP) where $p>q$, we find $\sigma_0^2(t)\sim
D_{ARAP}t$ for large $t$. The constants
$A_{SRAP}=2{\rho}^{-2}(p\mu_1/\pi)^{1/2}\mu_2/(\mu_1-\mu_2)$ and
$D_{ARAP}={\rho}^{-2}(p-q){\mu_1\mu_2}/{(\mu_1-\mu_2)}$ are
computed exactly. For the special case $q=0$ and $p=1$, $D_{ARAP}$ reduces
to $D_1$ computed earlier in Refs. \cite{KG,schutz}.
 \item We compute exactly the universal scaling function that
describes the crossover behavior of $\sigma_0^2(t)$ from the subdiffusive
$t^{1/2}$ growth to the diffusive $t$ growth as one switches on an
infinitesimal bias $(p-q)$.
 \item We generalize the single tracer particle fluctuation
$\sigma_0^2(t_0,t_0+t)$ to the fluctuation in the position difference of
two tagged particles defined as $\sigma_r^2(t_0,t_0+t)=\langle
[\zeta_{i+r}(t_0+t)-\zeta_i(t_0)]^2\rangle$. We show that in the steady
state $\sigma_r^2(t)=\lim_{t_0\to \infty}\sigma_r^2(t_0,t_0+t)$ grows
monotonically with $t$ for a fixed tag shift $r$ for the SRAP. For the
ARAP on the other hand, it grows with $t$ in a non-monotonic fashion with a
single minimum at a characteristic time $t^{*}=r/{\mu_1(p-q)}$.
 \item We also compute various scaling functions that describe the
crossover of the tracer fluctuations from their non-steady state behavior
to the steady state behavior as the waiting time $t_0\to \infty$.
 \end{enumerate}

The paper is organized as follows. In Sec. II, we define the model
precisely and set up our notations. In Sec. III, we calculate the equal
time correlation function for the RAP for all $p$ and $q$. Sec. IV
contains the exact calculation of the unequal time correlation function.
In Sec. V we compute the mean squared fluctuation in the displacement
of a single tracer particle. The Secs. VA and VB contain
respectively the discussions on the SRAP and the ARAP, while the crossover
between them is discussed in Sec. VC.  The Sec. VI contains the
generalization to the two-tag correlation functions. Finally we conclude
with a summary and discussion in Sec. VII.

\section{The Model and Preliminaries}

We consider a system of particles of average density $\rho$ located on a
real line. Let $x_i(t)$ denote the position of the $i$-th particle at time
$t$ (see Fig. 1). In an infinitesimal time interval $dt$, each particle
jumps with probability $pdt$ to the right, with probability $qdt$ to the
left and with probability $1-(p+q)dt$ it rests at its original location.
The actual amount by which a particle jumps (either to the right or to the
left) is a random fraction of the gap between the particle and its
neighboring particle (to the right or to the left). For example, the jump
to the right is by an amount $r_i^{+}(x_{i+1}-x_i)$ and to the left by
$r_i^{-}(x_i-x_{i-1})$. The random variables $r_i^{\pm}$ are independently
drawn from the interval $[0,1]$ and each is distributed according to the
same pdf $f(r)$ which is arbitrary. We start from an arbitrary but fixed
initial condition at $t=0$ and averaging of physical quantities is done
over all histories of evolution keeping the initial condition fixed. The
time evolution of the positions $x_i(t)$'s can be represented by the exact
Langevin equation
 \be
 x_i (t+dt)=x_i (t) + \gamma_i(t),
 \label{one}
 \ee
 where $\gamma_i (t)$ are random variables given by
 \be
 \gamma_i(t)=\cases
 {r_i^{+} (x_{i+ 1}(t)-x_i(t)) \quad\mbox{with prob $pdt$},\cr
 r_i^{-} (x_{i-1}(t)-x_i(t)) \quad\mbox{with prob $qdt$},\cr
 0 \mbox{\hskip16.2mm with prob $1-(p+q)dt$}.\cr}
 \label{two}
 \ee
 The random variables $r_i^{\pm}$ are independent and each is distributed
over the interval $[0,1]$ with the same pdf $f(r)$. The $k$-th moment of
the pdf is denoted by $\mu_k=\int_0^1 dr r^k f(r)$. Note that since $0\leq
r\leq 1$ and $f(r)\geq 0$, $\mu_1\geq \mu_2$.

We define a new random variable $\zeta_i(t)$ which measures the deviation
of $x_i(t)$ from its mean value
 \be
 \zeta_i (t)=x_i(t)-\langle x_i(t) \rangle.
 \label{three}
 \ee
 From Eqs. (\ref{one}) and (\ref{two}), one can easily derive the
evolution rules for the $\zeta_i$ variables. We find
 \be
 \zeta_i (t+dt)=\zeta_i (t) -(p-q) \frac{\mu_1}{\rho} dt +\eta_i(t),
 \label{four}
 \ee
 where $\eta_i(t)$ is given by
 \be
 \eta_i (t)=\cases
 {r_i^{+} (\zeta_{i+1}(t)-\zeta_i(t)+ {\rho}^{-1}) \quad \mbox{with prob
$pdt$},\cr
 r_i^{-} (\zeta_{i-1}(t)-\zeta_i(t)-{\rho}^{-1}) \quad \mbox{with prob
$qdt$},\cr
 0 \quad \mbox{\hskip21.2mm with prob $1-(p+q)dt$}.\cr}
 \label{five}
 \ee
 By definition, $\langle \zeta_i(t)\rangle =0$. Also from Eq.
(\ref{five}), it follows that $\langle \eta_i(t)\rangle= (p-q)
\frac{\mu_1}{\rho} dt$.
 
In this paper, we will focus on the mean squared displacement of a tagged
particle. It turns out that the asymptotic behavior of the mean squared
displacement depends crucially on whether one starts measuring these
fluctuations after some finite waiting time $t_0$ or if one first waits
for an infinite time and then starts measuring the statistics. The latter
corresponds to measuring the fluctuations in the steady state. This is
similar to the `approach to stationary' versus `stationary' regimes found
in various interface models\cite{pers}. This can be quantified precisely
in terms of the following correlation function,
 \bea
 \sigma_0^2(t_0,t_0+t) &=&\langle \left( \zeta_i(t+t_0)-\zeta_i(t_0)
\right)^2 \rangle,
 \label{six} \\
 &=& G_0(t\!+\!t_0)+ G_0(t_0) - 2 C_0(t_0, t_0\!+\!t)
 \label{seven},
 \eea
 where $G_r(t)=\langle \zeta_i(t)\zeta_{i+r}(t)\rangle$ is the equal time
correlation function and $C_r(t_0,t_0+t)=\langle
\zeta_i(t_0)\zeta_{i+r}(t_0+t)$ with $t>0$ denotes the unequal time
correlation function. For $t=0$, the unequal time correlation function
reduces to the equal time correlation function, $C_r(t_0,t_0)=G_r(t_0)$.
Note that we have assumed an infinite system so that the translational
invariance holds.  In the next two sections we calculate analytically the
correlation functions $G_r(t)$ and $C_r(t_0,t_0+t)$ respectively.

\section{Equal Time Correlation Function}

In this section, we calculate the equal time correlation function
$G_r(t)=\langle \zeta_i(t)\zeta_{i+r}(t)\rangle$ exactly for the RAP for
all $p$ and $q$. Our starting point is Eq. (\ref{four}) in conjunction
with Eq. (\ref{five}) describing the evolution of the $\zeta_i$ variables
with time. We consider the evolution equations (Eq. (\ref{four})) for both
$\zeta_i(t+dt)$ and $\zeta_{i+r}(t+dt)$, multiply them and then take the
average $\langle \rangle$ over all histories, keeping terms only upto $O(dt)$.
This yields, in the limit $dt\to 0$, the exact evolution equation of the
correlation function $G_r(t)$ and we obtain,
 \bea
 \frac{d}{dt} G_r(t)&=&\mu_1(p+q) \left[ G_{r+1}(t) +G_{r-1}(t)-2 G_r(t)\right]
\nonumber \\
 &+&\delta_{r, 0}\, \mu_2 (p+q) \left[ \rho^{-2} +2
\left(G_0(t)-G_1(t)\right)\right].
 \label{eight}
 \eea
 The Eq. (\ref{eight}) is valid for all positive and negative integers $r$
including $r=0$ and clearly $G_r(t)=G_{-r}(t)$. Thus the equation of
evolution for the two point correlations involve only two point correlations
and not higher order correlations. This closure property is crucial for
obtaining an exact solution for the correlation functions. The key reason
behind this closure lies in the fact that the random fractions
$r_i^{\pm}$'s at time $t$ are independent of the $\zeta_i(t)$. One
noteworthy fact about Eq. (\ref{eight}) is that the rates $p$ and $q$
make their appearance only as an overall multiplicative factor $(p+q)$.
We could absorb this factor into the time by doing a suitable rescaling, and
hence, the equal time correlation function
$G_r(t)$ is same for both the ARAP and the SRAP.

We note that this equation was also derived in Ref.\cite{schutz} by a
rather lengthy method, but was left unsolved. In this section, we derive
an exact solution of Eq. (\ref{eight}). Note that even though Eq.
(\ref{eight}) represents the diffusion equation (in discrete space) with a
source term at the origin $r=0$, its solution is nontrivial due to the
fact that the source term depends on $G_0(t)$ and $G_1(t)$ which need to
be determined self-consistently. Similar diffusion equations with source
term for the correlation functions have also appeared recently in the
context of aggregation models with injection\cite{RM2}. Before proceeding to
solve Eq. (\ref{eight}), we first set up our notations. We define the
standard Fourier transform
 \be
 {\bar G}(k,t)=\sum_{r=-\infty}^{\infty} G_r(t) e^{ikr},
 \label{fourier}
 \ee
the Laplace transform
 \be
 {\tilde G}_r(s)=\int_0^{\infty} G_r(t) e^{-st} dt,
 \label{laplace}
 \ee
and the joint Fourier-Laplace transform,
 \be
 F(k,s)=\int_0^{\infty}{\bar G}(k,t)e^{-st}dt=\sum_{r=-\infty}^{\infty}
{\tilde G}_r(s) e^{ikr}.
 \label{jointfl}
 \ee

Taking the joint Fourier-Laplace transform of Eq. (\ref{eight}) we obtain
 \be
 F(k,s)=\frac{\mu_2 (p+q) \left[ {\rho}^{-2}+2s\left({\tilde G}_0(s)-{\tilde
G}_1(s)\right)\right]}{s\left[s+2\mu_1 (p+q)(1-\cos k)\right]},
 \label{fl1}
 \ee
 where we have assumed that initially $G_r(0)=0$ which is true for any
fixed initial condition. For random initial condition, $F(k,s)$ will
contain additional terms arising from the initial condition, but one can
show that they do not contribute to the asymptotic large time properties
of $G_r(t)$ as long as the initial condition has only short ranged 
correlations.  We therefore use $G_r(0)=0$ without any loss of generality.

The Eq. (\ref{fl1}) contains two unknowns ${\tilde G}_0(s)$ and ${\tilde
G}_1(s)$. One of them, say ${\tilde G}_1(s)$ can however be expressed in
terms of ${\tilde G}_0(s)$ by taking directly the Laplace transform of Eq.
(\ref{eight}) for $r=0$ and using $G_1(t)=G_{-1}(t)$. This gives the
relation
 \be
 s{\tilde G}_0(s)=(p\!+\!q)\!\left[\frac{\mu_2 \rho^{-2}}{s}-2(\mu_1\!-\!\mu_2)
\left({\tilde G}_0(s)- {\tilde G}_1(s)\right)\right].
 \label{g0g1}
 \ee 
 Substituting Eq. (\ref{g0g1}) in Eq. (\ref{fl1}) we obtain
 \be
 F(k,s)=\frac{\mu_2}{(\mu_1-\mu_2)}{ {\left[
\mu_1 (p+q){\rho}^{-2}-s^2{\tilde G}_0(s)\right]}\over {s\left[ s+2\mu_1
(p+q)(1-\cos k)\right]}}.
 \label{fl2}
 \ee

We now have to determine ${\tilde G}_0(s)$ self-consistently. This can be
done by using the inverse Fourier transform
 \be
 {\tilde G}_r(s)= {1\over {2\pi}}\int_{-\pi}^{\pi} F(k,s)e^{-ikr}dk.
 \label{invg0}
 \ee
 Substituting the expression of $F(k,s)$ from Eq. (\ref{fl2}) in Eq.
(\ref{invg0}) at $r=0$, we obtain the exact ${\tilde G}_0(s)$
 \be
 {\tilde G}_0(s)={ {\mu_1\mu_2 (p+q)}\over {(\mu_1-\mu_2)}}{
{{\rho}^{-2}I(0,s)}\over {s\left[1+{ {\mu_2}\over
{(\mu_1-\mu_2)}}sI(0,s)\right]}},
 \label{g0s}
 \ee
 where $I(r,s)$ is given by the integral
 \bea
 I(r,s)&=&{1\over {2\pi}}\int_{-\pi}^{\pi} {{e^{-ikr}dk}\over {
\left[s+2\mu_1^\prime (1-\cos k)\right]} } \nonumber \\
&=&\frac{1}{\sqrt{s^2+4\mu_1^\prime  s}}
\!\left(\!\frac{2 \mu_1^\prime +s-\sqrt{s^2+4\mu_1^\prime  s}} {2
\mu_1^\prime }\right)^{|r|}\!\!,
 \label{irs}
 \eea
 where $\mu_1^\prime=\mu_1 (p+q)$.
Knowing ${\tilde G}_0(s)$ determines $F(k,s)$ completely by Eq.
(\ref{fl2}) and hence ${\tilde G}_r(s)$ for all $r$ by the Fourier
inversion formula in Eq. (\ref{invg0}).  We obtain
 \be
 {\tilde G}_r(s)={ {\mu_1\mu_2 (p+q)}\over {(\mu_1-\mu_2)} } {
{{\rho}^{-2}I(r,s)}\over { s\left[1+ { {\mu_2}\over {(\mu_1-\mu_2)} }
sI(0,s)\right]} },
 \label{grs}
 \ee
 where $I(r,s)$ is given by Eq. (\ref{irs}).

To obtain $G_r(t)$ we need to perform the inverse Laplace transform
$G_r(t)={\cal L}^{-1}[{\tilde G}_r(s)]$ with respect to $s$.  In general
for arbitrary $t$ this is difficult. However, for large $t$, this inverse
can be obtained in closed form. For large $t$, one needs to consider the
small $s$ behavior of ${\tilde G}_r(s)$ in Eq. (\ref{grs}).  Let us first
consider the case $r=0$. Putting $r=0$ in Eq. (\ref{irs}) and taking the
$s\to 0$ limit we find to leading order,
 \be
 I(0,s)\sim {1\over {2\sqrt{\mu_1 (p+q) s}} }.
 \label{i0s}
 \ee
 Substituting this small $s$ expression of $I(0,s)$ in Eq. (\ref{g0s}) and
taking the inverse Laplace transform we find that to leading order for
large $t$,
 \be
 G_0(t)=\frac{\sqrt{\mu_1(p+q)} \mu_2 \rho^{-2}}{(\mu_1-\mu_2) \sqrt{\pi}}
\sqrt{t}.
 \label{g0t1}
 \ee

Next we consider the behavior of $G_r(t)$ for $|r|>0$. From Eq. (\ref{irs})
it is clear that the appropriate scaling limit consists of taking
the limit $s\to 0$, $|r|\to \infty$ but
keeping $|r|\sqrt{s}$ fixed. In this scaling limit, 
Eq. (\ref{irs}) yields,
 \be
 I(r,s)= {1\over {2 \sqrt{\mu_1(p+q) s}}}\exp\left(\frac{-|r|\sqrt{s}}
{\sqrt{\mu_1 (p+q)}} \right).
 \label{irss}
 \ee
 We note that the formula for $I(r,s)$ in Eq. (\ref{irss}) reduces to Eq.
(\ref{i0s}) for $|r|=0$. This indicates that even though Eq. (\ref{irss})
was derived in the scaling limit, it continues to hold even for $r=0$.

Substituting this small $s$ expression of $I(r,s)$ in Eq. (\ref{grs}) and
taking the inverse Laplace transform we obtain for large $t$,
 \be
 G_r(t)= \frac{\sqrt{\mu_1 (p+q)} \mu_2 \rho^{-2}}{2(\mu_1-\mu_2)}{\cal
L}^{-1}\left[s^{-3/2}e^{-|r| \sqrt{s/[\mu_1(p+q)]}}\right].
 \label{grt1}
 \ee
 Fortunately the inverse Laplace transform in Eq. (\ref{grt1}) can be done
in closed form which gives us the following asymptotic scaling behavior of
the equal time correlation function $G_r(t)$,
 \be
 G_r(t)=\frac{\sqrt{\mu_1 (p+q)} \mu_2 \rho^{-2}}{(\mu_1-\mu_2) \sqrt{\pi}}
\sqrt{t} f_1\left(\frac{|r|}{2\sqrt{\mu_1 (p+q) t}}\right).
 \label{grt2}
 \ee
 Here $f_1(y)$ is a universal scaling function independent of the model
parameters such as $p$, $q$ and the moments $\mu_k$ of the pdf $f(r)$ and
is given by
 \be
 f_1(y)= e^{-y^2}-{\sqrt{\pi}}~y \mbox{~erfc}(y),
 \label{f1}
 \ee
 where ${\rm erfc}(y)=2/\sqrt{\pi}\int_y^{\infty}e^{-u^2}du$ is the
standard complimentary error function. This scaling function has the
asymptotic behaviors, $f_1(y)\sim 1- \sqrt{\pi}y$ as $y\to 0$ and $\sim
y^{-2}e^{-y^2}/2$ for $y\to \infty$.

As a final remark, we note again that if one puts $|r|=0$ in the formula
for $G_r(t)$ in Eq. (\ref{grt2}) one recovers the correct $G_0(t)$ as
given by Eq. (\ref{g0t1}). Thus the scaling range includes even the $r=0$
point. The Eq. (\ref{g0t1}) thus provides us the exact behavior of the
first two terms in the expression for $\sigma_0^2(t_0,t_0+t)$ in Eq.
(\ref{seven}). The remaining task is to evaluate the third term in Eq.
(\ref{seven}) which involves the unequal time correlation function and
this is done in the next section.

\section{Unequal time correlations}

In this section we compute the two time tag-tag correlation function
$C_r(t_0,t_0+t)=\langle \zeta_i(t_0)\zeta_{i+r}(t_0+t)\rangle$ for the
RAP.  We start at time $t_0$ and then evolve the $\zeta_{i+r}$ variables by
Eq. (\ref{four}) for all subsequent time. Let us first rewrite the Eq.
(\ref{four}) at time $t_0+t+dt$,
 \be
 \zeta_{i+r} (t_0\!+\!t\!+\!dt)=\zeta_{i+r} (t_0\!+\!t) -
(p\!-\!q) \frac{\mu_1}{\rho} dt +\!\eta_{i+r}(t_0\!+\!t).
 \label{four1}
 \ee
 We then multiply both sides of Eq. (\ref{four1}) by $\zeta_{i}(t_0)$
and average over the noise keeping terms only upto $O(dt)$. In the limit
$dt\to 0$, we obtain the exact evolution equation of the two time correlation
function,
 \bea
 \frac{dC_r(t_0,t_0\!+\!t)}{dt}&=&\mu_1\left[ p
C_{r+1}(t_0,t_0\!+\!t)+qC_{r-1}(t_0,t_0\!+\!t)\right. \nonumber \\
 &-& \left. (p+q)C_r(t_0,t_0+t)\right] \mbox{\hskip3mm for~} t \geq 0 .
 \label{crteqn}
 \eea
 Note that at $t=0$, the unequal time correlation function reduces to the
equal time correlation function $C_r(t_0,t_0)=G_r(t_0)$. Thus starting at
$t=0$ with the initial condition $C_r(t_0,t_0)=G_r(t_0)$, the function
$C_r(t_0,t_0+t)$ evolves with time $t$ according to the Eq.
(\ref{crteqn}). 

As in the preceding section, we define the Fourier transform ${\bar
C}(k,t_0,t_0+t)=\sum_{r=-\infty}^{\infty}C_r(t_0,t_0+t)e^{ikr}$. Taking
the Fourier transform of Eq.  (\ref{crteqn}) we obtain
 \be
 {\bar C}(k,t_0,t_0+t)={\bar G}(k,t_0)e^{-\mu_1 \alpha(k) t},
 \label{ckt}
 \ee
 where $\alpha(k)=p+q-(pe^{-ik}+qe^{ik})$ and ${\bar G}(k,t_0)$ is the
Fourier transform of the equal time correlation function as defined by Eq.
(\ref{fourier}). Taking further the Laplace transform
$H(k,s,t)=\int_0^{\infty}{\bar C}(k,t_0,t_0+t)e^{-st_0}dt_0$ of Eq.  
(\ref{ckt}) we obtain
 \be
 H(k,s,t)=F(k,s)e^{-\mu_1 \alpha(k) t},
 \label{hkst}
 \ee
 where $F(k,s)$ is given exactly by Eq. (\ref{fl2}) with ${\tilde G}_0(s)$
determined from Eq. (\ref{g0s}).

Proceeding as in the previous section, the Laplace transform ${\tilde
C}_r(s,t)= \int_0^{\infty}C_r(t_0, t_0+t)e^{-st_0}dt_0$ can then be
determined from the joint Fourier-Laplace transform $H(k,s,t)$ by the
inversion formula
 \be
 {\tilde C}_r(s,t)= {1\over {2\pi}}\int_{-\pi}^{\pi}H(k,s,t)e^{-ikr}dk,
 \label{crst1}
 \ee
 where $H(k,s,t)$ is given by Eq. (\ref{hkst}). Substituting in Eq.
(\ref{crst1}) the exact expression of $F(k,s)$ from Eq. (\ref{fl2})  and
that of ${\tilde G}_0(s)$ from Eq. (\ref{g0s}), we obtain the following
final expression of the Laplace transform
 \bea
 {\tilde C}_r(s,t)&=&{ {\mu_1\mu_2(p+q)}\over {(\mu_1-\mu_2)}}{
{{\rho}^{-2}}\over {s\left[1+{ {\mu_2}\over
{(\mu_1-\mu_2)}}sI(0,s)\right]}}\nonumber \\
 &\mbox{x}& \frac{1}{2\pi} \int_{-\pi}^{\pi} {{
e^{-ikr-\mu_1\alpha(k)t}dk}\over { \left[s+2\mu_1 (p+q)(1-\cos k)\right]} }.
 \label{crst2}
 \eea
 Note that for $t=0$, ${\tilde C}_r(s,t)$ as given by Eq. (\ref{crst2})
reduces to ${\tilde G}_r(s)$ given by Eq. (\ref{grs}) as expected. The
equation (\ref{crst2}) is central to our subsequent analysis for various
limiting behaviors.

\section{Mean Squared Tracer Auto-fluctuation}

In this section we calculate $\sigma_0^2(t_0,t_0+t)$ in the RAP using the
exact results for the equal time and two time correlation functions
obtained in the previous sections. We consider first the symmetric case
SRAP with $p=q$ in subsection A followed by the derivation for the
asymmetric case ARAP with $p>q$ in subsection B. In subsection C, we show
how the steady state fluctuation $\sigma_0^2(t)= lim_{t_0\to
\infty}\sigma_0^2(t_0,t_0+t)$ crosses over from the subdiffusive behavior
to the diffusive behavior as one switches on an infinitesimal bias and we
calculate the crossover scaling function exactly.

\subsection{SRAP}

Here we consider the symmetric case $p=q$. For the calculation of
$\sigma_0^2(t_0,t_0+t)$ we only need the asymptotic behavior of
$C_r(t_0,t_0+t)$ for $r=0$ as evident from Eq. (\ref{seven}). To obtain
$C_0(t_0,t_0+t)$ we need to invert the Laplace transform in Eq.
(\ref{crst2})  for $r=0$ and $p=q$. As before, this inversion is difficult
in general for all $t_0$. However the finite but large $t_0$ limit can be
worked out by analyzing the small $s$ behavior of Eq. (\ref{crst2}). It
turns out that the appropriate scaling limit in this case involves taking
$s\to 0$, $t\to \infty$ but keeping $st$ fixed. In this scaling limit, the
integration in Eq. (\ref{crst2})  can be carried out in closed form and we
obtain (with $p=q$),
 \be
 {\tilde C}_0(s,t)=\frac{\sqrt{2 p \mu_1} \mu_2
\rho^{-2}}{2(\mu_1-\mu_2)s^{3/2}}e^{st/2}{\rm erfc}\left(\sqrt{st/2}\right).
 \label{c0sts}
 \ee
 We then need to invert the Laplace transform in Eq. (\ref{c0sts}) with
respect to $s$ to obtain the asymptotic behavior of $C_0(t_0,t_0+t)$ for
large $t_0$. Fortunately this inversion can be done in closed form and we
obtain
 \be
 C_0(t_0,t_0+t)=\frac{\sqrt{2 p \mu_1} \mu_2
\rho^{-2}}{(\mu_1-\mu_2)\sqrt{\pi}}\sqrt{t_0}f_2\left({ {t}\over
{2 t_0}}\right),
 \label{c0t1}
 \ee
 where the scaling function $f_2(y)$ is again universal and is given by,
 \be
 f_2(y)=\sqrt{1+y}-\sqrt{y} .
 \label{sc2}
 \ee

We are now ready to compute $\sigma_0^2(t_0,t_0+t)$ from Eq.
(\ref{seven}). Using the result for the equal time correlation in Eq.
(\ref{g0t1}) and the one for the two time correlation in Eq. (\ref{c0t1}),
we obtain from Eq. (\ref{seven}) our main result
 \bea
 \lefteqn{ \sigma_0^2(t_0,t_0+t)=}\quad \nonumber \\
 & &\frac{\sqrt{2 p \mu_1} \mu_2 \rho^{-2}}{(\mu_1-\mu_2)\sqrt{\pi}}
\left[\sqrt{t_0+t}+\sqrt{t_0}-2\sqrt{t_0}f_2\left({ {t}\over
{2 t_0}}\right)\right],
 \label{smsf1}
 \eea
 where $f_2(y)$ is given by Eq. (\ref{sc2}). Note that this result in Eq.
(\ref{smsf1}) is derived in the scaling limit when both $t_0$ and $t$ are
large with their ratio $t/t_0$ kept fixed.

We now discuss two different limits of Eq. (\ref{smsf1}). First we
consider the steady state limit $t_0\to \infty$ with $t$ large but fixed.
In this limit, Eq. (\ref{smsf1}) yields
 \be
 \sigma_0^2(t)={\rm lim}_{t_0\to
\infty}\sigma_0^2(t_0,t_0+t)=\frac{2\sqrt{p\mu_1 } \mu_2
\rho^{-2}}{(\mu_1-\mu_2)\sqrt{\pi}}\sqrt{t}.
 \label{smsf2}
 \ee
 In the opposite limit, when the waiting time $t_0$ is finite (away from
the steady state) but the evolved time $t$ goes to infinity, we obtain from
Eq. (\ref{smsf1},
 \be
 {\rm lim}_{t\to \infty}\sigma_0^2(t_0,t_0+t)=\frac{\sqrt{ 2 p \mu_1 } \mu_2
\rho^{-2}}{(\mu_1-\mu_2)\sqrt{\pi}}\sqrt{t}.
 \label{smsf3}
 \ee
 Thus the mean squared auto-fluctuation in these two opposing limits differ
by a factor $\sqrt{2}$. The Eqs. (\ref{smsf1}), (\ref{smsf2}) and
(\ref{smsf3}) are amongst the important new results of this paper.

\subsection{ARAP}

In this subsection we calculate $\sigma_0^2(t_0,t_0+t)$ in the asymmetric
case when $p>q$. Once again we have to invert the Laplace transform in Eq.
(\ref{crst2}) for $r=0$ but now with $p>q$. In this case it turns out the
appropriate scaling limit consists of taking $s\to 0$, $t\to \infty$ as in
the SRAP but keeping $\sqrt{s} t$ instead of the scaling variable $st$ in
the SRAP. In this scaling limit, the integration in Eq. (\ref{crst2}) with
$r=0$ yields
 \be
 {\tilde C}_0(s,t)=\frac{\sqrt{\mu_1 (p+q)} \mu_2
\rho^{-2}}{2(\mu_1-\mu_2)s^{3/2}}e^{-(p-q)\sqrt{\mu_1 s/(p+q)}\,t}.
 \label{c0sta}
 \ee
 The Laplace transform in Eq. (\ref{c0sta}) can be inverted as in Eq.
(\ref{grt1}) and we obtain
 \be
 C_0(t_0,t_0+t)= \frac{\sqrt{\mu_1(p+q)} \mu_2
\rho^{-2}}{(\mu_1-\mu_2)\sqrt{\pi}}\sqrt{t_0}f_1\! \left[ {
{\sqrt{\mu_1}(p-q)t}\over {2\sqrt{(p+q) t_0}}}\right]\!,
 \label{c0t2}
 \ee
 where the universal scaling function $f_1(y)=e^{-y^2}-\sqrt{\pi}y~ {\rm
erfc}(y)$ is the same as in Eq.  (\ref{f1}).

Substituting the results in Eq. (\ref{c0t2}) and Eq. (\ref{g0t1}) in Eq.
(\ref{seven}) we obtain
 \bea
 \lefteqn{\sigma_0^2(t_0,t_0+t)= \frac{\sqrt{\mu_1 (p+q)} \mu_2
\rho^{-2}}{(\mu_1-\mu_2)\sqrt{\pi}}}\qquad\nonumber \\
 &&\mbox{x}\left[\sqrt{t_0+t}+\sqrt{t_0}-2\sqrt{t_0}f_1 \left( {
{\sqrt{\mu_1}(p-q)t}\over {2\sqrt{(p+q)t_0}}}\right)\right].
 \label{amsf1}
 \eea

As in the SRAP we now discuss the two different limits. In the steady
state $t_0\to \infty$ with fixed large $t$ we obtain from Eq. (\ref{amsf1}),
 \be
 \sigma_0^2(t)={\rm lim}_{t_0\to \infty}\sigma_0^2(t_0,t_0+t)=\frac{ \mu_1
\mu_2 \rho^{-2}(p-q)}{(\mu_1-\mu_2)}t.
 \label{amsf2}
 \ee
 Thus in this case $\sigma_0^2(t)$ grows diffusively for large $t$,
$\sigma_0^2(t)=D_{ARAP}t$ where the diffusion constant,
 \be
 D_{ARAP}={\rho}^{-2}(p-q){{\mu_1\mu_2}\over {(\mu_1-\mu_2)}},
 \label{dc}
 \ee
 depends explicitly on $p$ and $q$.  For $q=0$ and $p=1$, it reduces to
the expression $D_1={\rho}^{-2}\mu_1\mu_2/(\mu_1-\mu_2)$ derived by Krug
and Garcia using the independent jump approximation\cite{KG} and later
rederived by Sch\"utz\cite{schutz} using a different approach.

We make a brief comment here on the approach used in Ref.\cite{schutz} in
deriving the diffusion constant $D_1$. In his approach, Sch\"utz started
with the evolution equation (\ref{eight}) for the equal time correlation
function and then used a chain of arguments to derive the diffusion
constant $D_1$. His approach didn't require any knowledge of the two time
correlation function or even the solution of the equal time correlation
function. As evident from the definition in Eq. (\ref{seven}) that
$\sigma_0^2(t_0,t_0+t)$ requires the knowledge of both the equal and the
two time correlation functions. Thus it was rather remarkable that the
correct value of the diffusion constant for $q=0$ and $p=1$ was recovered
in Ref. \cite{schutz}. However this turns out to be purely fortuitous.
Note that the evolution equation (\ref{eight}) is independent of the bias in
the system.
Thus the approach of Sch\"utz would predict that the diffusion
constant is also completely independent of the bias $(p-q)$ and is always given
by $D_1$ (provided $t$ is scaled by $(p+q)$). 
This is clearly wrong as evident from the exact expression in
Eq. (\ref{dc}). In particular for the symmetric case $p=q=1/2$, the arguments
of Ref. \cite{schutz} would predict a diffusive growth of $\sigma_0^2(t)$
with the diffusion constant $D_1$. This is again incorrect since for $p=q$
the diffusion constant is $0$ from Eq. (\ref{dc}) which is consistent with
the correct asymptotic subdiffusive growth of $\sigma_0^2(t)$ as given
exactly by Eq. (\ref{smsf2}). The problem in the derivation of Sch\"utz
can be traced back to the fact that his arguments only used equal time
correlations (which involve only $(p+q)$) and not the two time
correlations. The dependence on the bias $(p-q)$ of the diffusion constant
$D_{ARAP}$ comes only from the two time correlations. The derivation of
Ref.\cite{schutz} misses this important fact and is rather fortuitous to
obtain the correct value $D_1$ of the diffusion constant for the special
case when $p=1$ and $q=0$.
    
We end this subsection by discussing the other limit when the system is
away from the steady state, i.e. when $t_0$ is large but finite and $t\to
\infty$. In this limit, we obtain from Eq. (\ref{amsf1})
 \be
 {\rm lim}_{t\to \infty}\sigma_0^2(t_0,t_0+t)=\frac{\sqrt{\mu_1 (p+q) } \mu_2
\rho^{-2}}{(\mu_1-\mu_2)\sqrt{\pi}}\sqrt{t},
 \label{amsf3}
 \ee
 the same result as in the SRAP in this limit [Eq. (\ref{smsf3})]. Thus
away from the steady state the tracer particle doesn't sense the presence
of bias. The exact result in Eq. (\ref{amsf3}) is consistent with that of
Krug and Garcia using a phenomenological hydrodynamic equation\cite{KG}.

\subsection{Crossover Between SRAP and ARAP}

In the previous subsections, we have seen that the asymptotic large $t$
behavior of $\sigma_0^2(t_0,t_0+t)$ does not depend on the bias $(p-q)$, when
the system is away from the steady state (finite $t_0$). However, in the
steady state ($t_0\to \infty$) it behaves rather differently in the
symmetric and asymmetric cases. In the steady state of the SRAP ($p=q$),
$\sigma_0^2(t)\sim t^{1/2}$ while for the ARAP ($p>q$), $\sigma_0^2(t)\sim
t$. Thus a natural question is: How does the behavior of $\sigma_0^2(t)$
crosses over from the subdiffusive growth for $p=q$ to the diffusive
growth as one switches on an infinitesimal bias $(p-q)$? In this
subsection we compute exactly the universal scaling function that
describes this crossover behavior of $\sigma_0^2(t)$.

To calculate the crossover behavior we return to our central equation
(\ref{crst2}) with $r=0$. We have seen in the previous subsections that in
the scaling limit $s\to 0$ and $t\to \infty$ of the Eq. (\ref{crst2}), the
appropriate scaling variable that is kept fixed is $st$ for $p=q$, where
as, it is $\sqrt{s}t$ for $p>q$. Thus, to compute the crossover behavior,
we need to keep the leading order terms in both of these scaling variables
fixed while expanding Eq. (\ref{crst2}) for small $s$ and large $t$. This
makes the calculation of the crossover behavior somewhat delicate. To
leading order, we find after elementary algebra
 \bea
 \lefteqn{ {\tilde C}_0(s,t)=
 \frac{\sqrt{\mu_1(p+q)} \mu_2 \rho^{-2}}{(\mu_1-\mu_2)s^{3/2}}}\qquad
\nonumber\\
&&\mbox{x}{1\over
{2\pi}}\int_{-\infty}^{\infty} { {
e^{-i(p-q)\sqrt{\mu_1s/(p+q)}\,tz-stz^2/2}}\over {1+z^2}}dz.
 \label{crst3}
 \eea
 Note that for the symmetric case $p=q$, the integral in Eq. (\ref{crst3})
can be done and we get back Eq. (\ref{c0sts}) of Sec. VA. Similarly,
for the asymmetric case $p>q$, in the limit $s\to 0$ keeping the scaling
variable $\sqrt{s}t$ fixed, one drops the second term in the exponential
in the integrand of Eq. (\ref{crst3}) and performing the resulting
integral we recover the Eq. (\ref{c0sta}) of Sec. VB.

To compute the crossover behavior we need to keep both the terms inside
the exponential in the integrand of Eq. (\ref{crst3}) and perform the
integral. Fortunately this integral can be done in closed form  using the
standard convolution theorem. We omit the details here and present only
the final result,
 \bea
 \lefteqn{ {\tilde C}_0(s,t)= \frac{\sqrt{\mu_1 (p+q)} \mu_2
\rho^{-2}}{4(\mu_1-\mu_2)s^{3/2}}}\qquad\nonumber\\
 &&\mbox{x}\left[e^{u-v}{\rm erfc}\left( {{2u-v}\over {2\sqrt{u}} }\right)
+e^{u+v}{\rm erfc}\left( {{2u+v}\over {2\sqrt{u}} }\right)\right],
 \label{cr1}
 \eea
 where $u=st/2$ and $v=(p-q)\sqrt{\mu_1 s/(p+q)}\,t$. We then expand the Eq.
(\ref{cr1}) further for small $s$ to obtain the steady state $t_0\to
\infty$ behavior. Note that we needed to first do the integral in Eq.
(\ref{crst3}) and then take the $s\to 0$ limit. The reverse order
unfortunately does not work.  Expanding Eq. (\ref{cr1}) for small $s$,
keeping only the leading order terms in $s$ and finally inverting the
Laplace transform of the resulting expression we obtain for large $t_0$
 \bea
 \lefteqn{C_0(t_0,t_0+t)=\frac{\sqrt{\mu_1(p+q)} \mu_2
\rho^{-2}}{(\mu_1-\mu_2)\sqrt{\pi}}} \quad\nonumber \\
 &&\mbox{x}\left[\sqrt{t_0}-\sqrt{ {{t}\over {2}}}e^{-w^2(t)}
 -{{\sqrt{\pi \mu_1}(p-q)t}\over {2\sqrt{p+q}}}{\rm
erf}\left(w(t)\right) \right]\!,
 \label{cr2}
 \eea
 where $w(t)=(p-q)\sqrt{\mu_1 t/ \left[2(p+q)\right]}$.

We now use the results from Eq. (\ref{cr2}) and Eq. (\ref{g0t1}) in Eq.
(\ref{seven})  and eventually take the strict $t_0\to \infty$ limit to
obtain the final form of the steady state auto-fluctuation
 \bea
 \sigma_0^2(t)&=&{\rm lim}_{t_0\to \infty}\sigma_0^2(t_0,t_0+t)
\nonumber\\
 &=&\frac{ \mu_1 \mu_2 (p-q)\rho^{-2}}{(\mu_1-\mu_2)}t Y\left[ {
{(p-q)\sqrt{\mu_1 t}}\over {\sqrt{2(p+q)}}}\right],
 \label{cr3}
 \eea
 where $Y(y)$ is a universal crossover scaling function given by
 \be
 Y(y)={\rm erf}(y)+ {1\over {\sqrt {\pi}}}{e^{-y^2}\over {y}}.
 \label{crsc}
 \ee
 The scaling function has the asymptotic behavior $Y(y)\sim
1/(\sqrt{\pi}y)$ as $y\to 0$ and $Y(y)\to 1$ and $y\to \infty$. Note that
for fixed $p-q>0$, if we take the limit $t\to \infty$ in Eq. (\ref{cr3})
(which corresponds to $y\to\infty$ in the scaling function in Eq.
(\ref{crsc})) we recover the result of Eq. (\ref{amsf2}). Similarly if we
take the $p-q\to 0$ limit for fixed $t$ in Eq. (\ref{cr3}) (corresponding
to taking $y\to 0$ limit in the scaling function $Y(y)$), we recover, as
expected, the result of Eq. (\ref{smsf2}) of the symmetric case. Thus the
Eq. (\ref{cr3}) and the associated scaling function $Y(y)$ in Eq.
(\ref{crsc}) describes the crossover behavior from the subdiffusive to
diffusive growth as one switches on an infinitesimal bias.

\section{Generalization to the two-tag correlation Function}

So far in this paper we have concentrated only on the mean squared
auto-fluctuation of a tracer particle, $\sigma_0^2(t_0,t_0+t)= \langle
\left( \zeta_i(t+t_0)-\zeta_i(t_0) \right)^2 \rangle$. A natural
generalization of the auto-fluctuation would be to study the two-tag
correlation function defined as
 \bea
 \sigma_r^2(t_0,t_0+t) &=&\langle \left( \zeta_{i+r}(t+t_0)-\zeta_{i}(t_0)
\right)^2 \rangle,
 \label{tt1} \\
 &=& G_0(t\!+\!t_0)+ G_0(t_0) - 2 C_r(t_0, t_0\!+\!t)
 \label{tt2},
 \eea
 where $G_r(t)$ and $C_r(t_0,t_0+t)$ are the usual equal time time and the
two time correlation functions already defined and derived in the previous
sections. Note that for $r=0$, the two-tag correlation in Eq. (\ref{tt1})
reduces to the single tag function $\sigma_0^2(t_0,t_0+t)$.

Of particular interest would be to compute the two-tag correlation
function in the steady state, i.e. $\sigma_r^2(t)=\lim_{t_0\to
\infty}\sigma_r^2(t_0,t_0+t)$. For the exclusion process this two-tag
correlation function was first introduced in Ref. \cite{MB2} and the
presence of bias was found to have a dramatic effect on the time
dependence of $\sigma_r^2(t)$ for a fixed $r$. It was found numerically
that while in the SEP $\sigma_r^2(t)$ increases monotonically with $t$ for
a fixed tag-shift $r$, in the ASEP $\sigma_r^2(t)$ has a non-monotonic
dependence on $t$\cite{MB2}. In the ASEP $\sigma_r^2(t)$ first decreases
with time $t$, becomes a minimum at some characteristic time $t^{*}$ and
then starts increasing again. A harmonic model was proposed in Ref.
\cite{MB2} for which $\sigma_r^2(t)$ could be computed analytically and
was found to be in qualitative agreement with the numerical results of the
exclusion process. But to the best of our knowledge, exact calculation of
$\sigma_r^2(t)$ for the exclusion process is still an unsolved problem.
However it turns out that for the RAP, it is possible to compute this
function $\sigma_r^2(t)$ exactly for large $t$. The exact solution of
$\sigma_r^2(t)$ in the RAP, as shown below, shares the similar features as
in the exclusion process.

From Eq. (\ref{tt2}) it is evident that we just need to compute the large
$t_0$ behavior of the two time correlation function $C_r(t_0,t_0+t)$ for
fixed nonzero $r$. In the previous sections we have analyzed in detail the
$r=0$ case. It turns out that the analysis for $r\neq 0$ proceeds more or
less in the same manner as in the $r=0$ case. We start, once again, from
the central equation (\ref{crst2}). To avoid separate calculations for the
SRAP and the ARAP, we take the line of approach used to calculate the
crossover behavior in subsection VC. For $r\neq 0$, it turns out that the
equation (\ref{crst3}) gets replaced by a similar looking equation,
 \bea
 \lefteqn{ {\tilde C}_r(s,t)=
\frac{\sqrt{\mu_1(p+q)} \mu_2 \rho^{-2}}{(\mu_1-\mu_2)s^{3/2}}}\qquad \nonumber\\
 &&\mbox{x}{1\over
{2\pi}}\int_{-\infty}^{\infty} { {
e^{-iz\sqrt{s/[\mu_1(p+q)]}R-stz^2/2}}\over {1+z^2}}dz,
 \label{tt3}
 \eea
 where $R=r+\mu_1(p-q)t$ signifies the drift of the particles to the right
with average velocity $\mu_1(p-q)$ for $p>q$. Clearly for $r=0$, Eq.
(\ref{tt3}) reduces to Eq. (\ref{crst3}). Starting with Eq. (\ref{tt3}) we
then follow exactly the same steps as used in subsection VC. Since the
steps are identical we skip all the details and present only the final
result. In the strict steady state limit $t_0\to \infty$, we finally obtain
the following scaling form
 \bea
 \sigma_r^2(t)&=&{\rm lim}_{t_0\to \infty}\sigma_r^2(t_0,t_0+t)\nonumber
\\
 &=& \frac{\sqrt{2\mu_1(p\!+\!q)}\mu_2
\rho^{-2}}{(\mu_1-\mu_2)\sqrt{\pi}}\sqrt{t}W\!\left( {R\over
{\sqrt{2\mu_1(p+q)t}}}\right)\!,
 \label{tt4}
 \eea
 where $R=r+\mu_1(p-q)t$ and $W(y)$ is again a universal scaling function
given by,
 \be
 W(y)=e^{-y^2}+\sqrt{\pi}y\,{\rm erf}(y).
 \label{tt5}
 \ee
Clearly $\mu_1(p-q)t$ represents the average drift while
$l(t)=\sqrt{2\mu_1(p+q)t}$ represents the diffusive length scale.

We note that the scaling function $W(y)$ is a symmetric function of $y$
about $y=0$ with a minimum at $y=0$. For the SRAP, $p=q$ and hence $R=r$.
Thus for a fixed $r$, it follows from Eq. (\ref{tt4}) that $\sigma_r^2(t)$
increases monotonically with $t$. For the ARAP on the other hand, $p>q$
and $R=r+\mu_1(p-q)t$. If one fixes $r$ to a negative value and increases
$t$, the variable $R$ remains negative till the characteristic time
$t=t^*=r/\mu_1(p-q)$, beyond which it becomes positive. The scaling
variable $y=R/\sqrt{2\mu_1(p+q)t}$ behaves in the same way. Thus
$\sigma_r^2(t)$ in Eq. (\ref{tt4}) first decreases with time, becomes a
minimum at $t=t^*=-r/\mu_1(p-q)$ and then starts increasing again.  In Fig.
(2) we plot the function $\sigma_r^2(t)$ in Eq. (\ref{tt4}) for both the
SRAP (with $p=q=1/2$) and the ARAP (with $p=1$ and $q=0$) for the same
value of $r=-2$ and choosing the parameter values $\mu_1=1/2$, $\mu_2=1/4$,
$\rho=1$. These features in the RAP, derived here exactly, are
qualitatively similar to those in the exclusion process studied in Ref.
\cite{MB2}.
\begin{figure}
\narrowtext\centerline{\epsfxsize\columnwidth \epsfbox{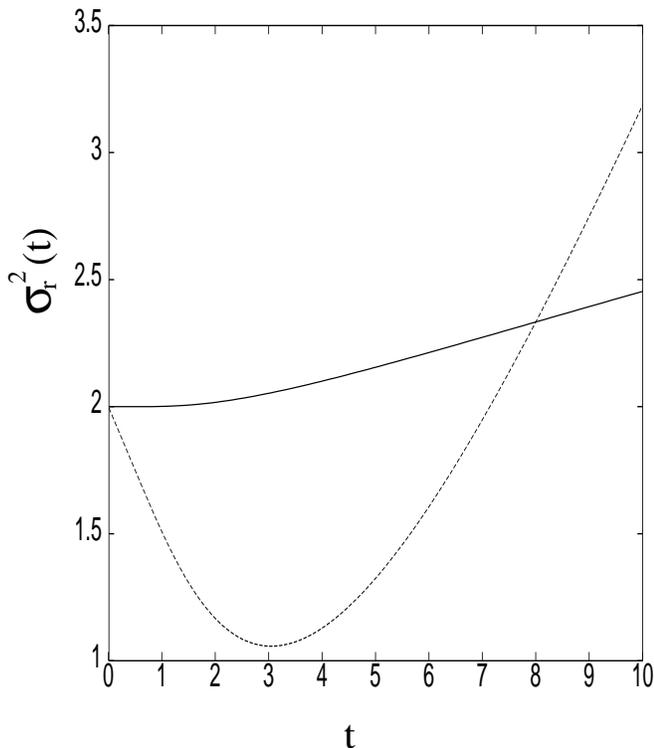}}
\caption{The steady state two-tag correlation function $\sigma_r^2(t)$ in
Eq. (\ref{tt4}) plotted as a function of $t$ for fixed $r=-2$ for parameter
values $\mu_1=1/2$, $\mu_2=1/4$ and $\rho=1$. The solid line shows the
monotonic growth of $\sigma_r^2(t)$ with $t$ for the SRAP ($p=q=1/2$)
while the dashed line shows the non-monotonic growth for the ARAP ($p=1$
and $q=0$).}
\end{figure}  

\section{Conclusions} 
 In this paper we have studied analytically the mean squared fluctuations
in the diffusion of both a single tagged particle and two tagged particles
in the random average process (RAP) for all values of the hopping rates
$p$ and $q$ in one dimension. We have shown that in the steady state, the
auto-fluctuation of a tagged particle grows subdiffusively as
$\sigma_0^2(t)\sim A_{SRAP}t^{1/2}$ for $p=q$ and diffusively
$\sigma_0^2(t)\sim D_{ARAP}t$ for $p>q$ where
$A_{SRAP}=2{\rho}^{-2}(p\mu_1/\pi)^{1/2}\mu_2/(\mu_1-\mu_2)$ and
$D_{ARAP}={\rho}^{-2}(p-q){\mu_1\mu_2}/{(\mu_1-\mu_2)}$. These behaviors
of $\sigma_0^2(t)$ are similar to those in the simple exclusion process
except the prefactors
$A=(2/\pi)^{1/2}(1-\rho)/\rho$\cite{harris,arratia,AP} and
$D=(p-q)(1-\rho)$\cite{DF,KvB} are different in the exclusion process.
Besides the steady state mean squared two-tag fluctuation $\sigma_r^2(t)$
in the RAP grows monotonically with $t$ for $p=q$ and non-monotonically
for $p>q$, in much the same way as in the exclusion process.

These findings raise the question whether or not the RAP is in the same
universality class as the simple exclusion process in one dimension.
Perhaps the RAP is just a coarse grained version of the exclusion process
in one dimension? The answer to this question seems to be in the negative
due to a very crucial difference between the two processes. In the
exclusion process for $p>q$, it is well known that there exists an
anomalous $t^{2/3}$ growth hidden in the problem apart from the usual
$t^{1/2}$ and $t$ growth\cite{vBKS,MB1}. This anomalous growth shows up
either in the mean squared fluctuation of the center of mass of the
particles when viewed from a special moving frame\cite{vBKS} or
alternately in the two-tag correlation function $\sigma_r^2(t)$ if one
chooses the tag shift $r$ to be sliding with time with a special velocity
$r=-\rho^2(p-q)$\cite{MB1}. It turns out that the prefactor of this
$t^{2/3}$ growth is proportional to $\propto {{d^2j(\rho)}\over
{d\rho^2}}$ where $j(\rho)$ is the current density in a hydrodynamical
description\cite{MB1}. For the exclusion process, $j(\rho)=\rho(1-\rho)$
and hence the prefactor is nonzero. For the RAP on the other hand,
$j(\rho)=\mu_1(p-q)$ and is independent of $\rho$. This is because
$j(\rho)=\rho \langle v\rangle$ where the average velocity $\langle
v\rangle=\mu_1(p-q)/\rho$ as can be easily derived from Eqs. (\ref{one})
and (\ref{two}). As a result, for the RAP, the anomalous $t^{2/3}$ growth
is absent which puts it in a different universality class than the simple
exclusion process. In this sense the RAP seems to be closer to the
harmonic model studied in Ref. \cite{MB2}.

In this paper we have considered the RAP only in one dimension. An obvious
generalization would be to higher dimensions. A natural way to generalize
the model to higher dimensions would be as follows. One considers
particles located in the continuous $d$-dimensional space. In a small time
interval $dt$, each particle makes a list of all its nearest neighbors in
various directions in space, chooses one of them at random and jumps in
the corresponding direction by a random fraction of the Euclidean distance
to that neighbor. This is an isotropic version, a generalization of the
SRAP. Similarly one can define an anisotropic version as well. To the best
of our knowledge, the RAP has not been studied so far in higher
dimensions. The question of tracer diffusion in higher dimensions,
especially in two dimensions where one may expect a logarithmic
correction, also remains completely open.

\end{multicols}

\end{document}